# General constraints on influential error sources for super-high accuracy star tracker

J. Zhang[*], Y. C. Hao, L. Wang, Y. Long

*Beijing Institute of Control and Engineering, 16 Zhongguancun Nansanjie, Haidian district, Beijing 100190, China*
*\*Corresponding author:* zhangjun208@mails.ucas.ac.cn

Though in-orbit calibration is adopted to reduce position error of individual star spot down to 0.02pixel on star tracker, little study has been conducted on the accuracy to what extent for some significant error sources which often leads to in-orbit correction inefficiency. This study presents the general theory and estimates of the minimum error constraints, including not only on position but also on intensity and scale of Gaussian shaped profile based on Cramer Rao Lower Bound(CRLB) theory. By imposing those constraints on motion, drift in focal length and so on, margins of in-flight error sources and the final accuracy of star tracker can be analytically determined before launch.

A star tracker serves as the highest precision sensor providing absolute attitude information to spacecraft after star extraction and identification processes. Many spacecraft for advanced space missions require star tracker 1 pointing accuracy down to 0.2 arcsec. This indicates position error of individual star spot should be lower than 0.02pixel. However, various influential error sources, such as device noises, optical aberrations, CCD photo response non uniformity(PRNU), temperature-dependent change in focal length, motion, velocity aberration and jitter of motion, can cause a position error much bigger than 0.02pixel[1-5]. The position errors are consisted of spatial domain error, time domain error, bore-sight alignment error and NEA error[1-4]. Different error feathers the distinct calibration process. Although all super-high accuracy star trackers have implemented in-orbit calibration technology to reduce spatial domain error and bore-sight alignment error[1-3], little research has been conducted on the accuracy to what extent for the other error sources both in theoretical aspect and industrial design. For example, without knowledge of error margins for some influential factors, the error reduction level of them is unclear leading to in-orbit correction inefficiency.

In order to reduce position error to a certain point, we present a general theory and estimates of error constraints on influential error sources for in-orbit calibration. We exploit the minimum variance errors not only of position but also of intensity and scale of Gaussian shaped star spot based on Cramer Rao Lower Bound(CRLB) theory[6-12]. In the virtue of simplicity, efficiency and integrality, those error indicators are then proposed as the total constraints on motion, jitter of motion, drift in focal length and other influential error sources. The error budgets of them are ultimately derived, which also make clear the mechanisms of various conventional technologies in industrial community and provide new potential solution to unsolved problems.

The **first** and foremost variance error comes from device noises[6,10-12]. In the positioning procedure for star spot generated in optoelectronic devices under static condition, device noises are random variables with no possibility to be subtracted. In statistics, CRLB is used to estimate such locating variance error. Lindegren and Winick had obtained CRLB on the performance for positioning estimator on CCD device[6,10]. In Winick's paper, CCD array is assumed to be consisted of identical discrete square pixels without dead space and star spot is represented by Gaussian shaped profile. As illustrated in literature[6,9-13], the minimum position error for device under static condition can be defined as square root of CRLB and given by

$$\xi_x \triangleq \sqrt{CRLB(x_c)} = \sqrt{2}\sigma_s/R \quad (1)$$

where $\sigma_s$ denotes the width of Gaussian signal with $\sigma_s = \beta L, L = \lambda f/\pi D$. $\lambda$ denotes the center wavelength of star spectrum. $f$ denotes the real focal length. $D$ denotes the aperture size of incident lens and $\beta$ is the ratio of the width of defocus star spot to Airy spot[4]. $R$ represents signal to noise ratio(SNR).

Eq.(1) indicates an error on scale and intensity can cause additional error on position. Therefore, the minimum square root errors of $I_0$ and $\sigma_s^2$ should be derived. Take the derivative with respect to $I_0$ and $\sigma_s^2$, and It yields

$$\xi_{I_0} \triangleq \sqrt{CRLB(I_0)} = \frac{I_0}{R}, \xi_{\sigma^2} \triangleq \sqrt{CRLB(\sigma^2)} = \frac{4\sigma_s^2}{\sqrt{3}R} \quad (2)$$

For Gaussian shaped signal, once the accuracies of intensity, scale and position are guaranteed to be as lower as CRLB, no gains will be further obtained under a better estimator. As illustrated in Eq.(1) and Eq.(2), any growth of bias or variance error on intensity, scale and position will lead to the final position accuracy beyond CRLB. Since the position error is proportional to $\sigma_s/R$, $\xi_{I_0}$, $\xi_x$ and $\xi_{\sigma^2}$ would be the full constraints to limit some influential error sources. In this sense, we propose CRLBs in Eq.(1) and Eq.(2) as the baseline for any other influential factor on star tracker. Then the error constraints on those parameters could be defined as

$$\delta\Phi \leq \zeta_\Phi \xi_\Phi, \zeta_\Phi < \infty, \Phi \in \{x, \sigma^2, I_0\} \quad (3)$$

Once Eq.(3) has been admitted, It means that the acquired star data set can not only output the attitude, but also to make on-line calibration directly.

For in-flight star tracker, in addition to device noises, the motion of platform, the drift in focal length, the range of star magnitude and color dispersion of stars are all influential error sources contributing to the final accuracy[1-4]. In order to obtain the concrete error budget, models on those factors should be sought and derived in details. After that, the error constraints in Eq.(3) are put on those factors and in-depth discusses are also given.

Spacecraft motion is the **second** large error source. Due to the large angular velocity of the spacecraft, the star spot will be smeared during the exposure time and ultimately forms obvious trails[2-4,14-16]. This will affect star detection sensitivity and position accuracy seriously. To derive the CRLB error on motion is an elaborate and elegant process.

Following to the "Digital Image Processing" written by Gonzalez and Woods[16], the pattern of a moving star spot could be defined as

$$\tilde{S}(x,T) = \int_0^T \frac{I_0}{T\sqrt{2\pi}\sigma_s} \exp\left[\frac{-(x-x_c(t))^2}{2\sigma_s^2}\right] dt \quad (4)$$

where $T$ is the total exposure time and $x_c(t)$ represents the position of star spot at time $t$.

To obtain accurate $x_c(t)$, attitude matrix $C(t)$ should be derived. According to literatures[14-15], If the 2nd order derivative of attitude matrix transform exists at initial time $t$ under maneuvering condition, the Taylor expansion of $C(t+\Delta t)$ can be given by

$$C(t+\Delta t) = [I - \omega_t^\times \Delta t + \frac{1}{2}(\omega_t' \omega_t' - \omega_t' \omega_t I - \dot{\omega}_t^\times)(\Delta t)^2]C(t) \quad (5)$$

where $\omega_t' = \begin{bmatrix} \omega_{xt} & \omega_{yt} & \omega_{zt} \end{bmatrix}$. $\omega_t^\times$ denotes the cross product matrix to the angular velocity vector $\omega_t$. $\Delta t$ denotes the temporal integration time.

Assume Y and Z axes are stationary and the shift of position satisfies $\Delta x_c(t) \ll f$ [15]. Note that focal length and angular rate drift all the time. From Eq.(5), the position and velocity of the moving star spot can be modeled as

$$x_c(t) = x_0 + vt + 0.5at^2, v = f\omega_{yt}, a = f\dot{\omega}_{yt} - \omega_{yt}^2 x_c(t) \quad (6a)$$

$$v = v_0 + \varsigma_{\omega v} + \varsigma_{fv}, v_0 = \bar{f}\bar{\omega}_{yt}, \varsigma_{\omega v} = \bar{f}\delta\omega_{yt}, \varsigma_{fv} = \delta f\bar{\omega}_{yt} \quad (6b)$$

where $x_0$ is the initial position at time $t$, $a$ and $v$ denote acceleration and velocity with units of pixel/s² and pixel/s on CCD focal plane, respectively. $\varsigma_{\omega v}$ represents jitter of motion with distribution $N(0, \sigma_{\omega v}^2)$. $\varsigma_{fv}$ represents velocity aberration[1]. $\bar{f}$ denotes the average of $f$. $\delta f$ denotes the drift from $f$. $\bar{\omega}_{yt}$ denotes the mean value of angular rate $\omega_{yt}$ and $\delta\omega_{yt}$ denotes the jitter of $\omega_{yt}$.

From Eq.(6a) and Eq.(6b), when $x(t)\omega_{yt} t > 2f$ or $\Delta\omega_{yt} > 2\omega_{yt}$, It indicates that a bias error from acceleration can emerge in the region far from the center of FOV under high angular velocity or at the time when control moment is exerted.

To obtain accurate CRLB errors on position and scale under acceleration, a novel scheme is proposed to derive the motion compensation. The novel scheme treats that the whole $T$ is consisted of infinite intervals $T_m$ with $\sum_m T_m = T$, $T_m \propto v^{-1}$. In each $T_m$, the angular rate is deemed as constant and the star spot is thought to follow ideal Gaussian shaped pattern. As $T_m$ leaps into $T_{m+1}$, $x_m$ can be transit into $x_{m+1}$ immediately. The time located at $x_m$ is propositional to $1/v_m$. Then the center of star spot can be given by

$$x_c = E\left[\left(\int_0^T \frac{1}{v} dt\right)^{-1} \int_0^T \frac{x_0 + vt + 0.5at^2}{v} dt\right] - x_0$$

$$= \frac{1}{4a} \frac{v_1^2 + \sigma_{\omega v}^2 - v_0^2}{\ln v_1 - \ln v_0} - \frac{v_0^2}{2a}, v_1 = v_0 + aT + \varsigma_{fv} \quad (7)$$

Note that for the case $v = 0, a \neq 0$, $x_c \approx 0.125aT^2$.

The final scale of the smeared star spot could be derived based on literature[9,12]

$$\sigma_s'^2 \triangleq \sigma_s^2 + \Delta\sigma_{sv}^2 \quad (8)$$

where

$$\Delta\sigma_{sv}^2 = (\Delta x')^2 - \frac{\Delta x'(v_1^2 + \sigma_{\omega v}^2 - v_0^2)}{2a(\ln v_1 - \ln v_0)} + \frac{v_1^4 + \sigma_{\omega v}^4 + 6v_1^2 \sigma_{\omega v}^2 - v_0^4}{16a^2(\ln v_1 - \ln v_0)}$$

$$\Delta x' = \frac{1}{4a} \frac{v_1^2 + \sigma_{\omega v}^2 - v_0^2}{\ln v_1 - \ln v_0}$$

Note that for the case $v = 0, a \neq 0$, $\Delta\sigma_{sv}^2 \approx 0.015625 a^2 T^4$.

According to Eq.(1), an approximate CRLB for the moving star spot under acceleration is obtained

$$\xi_{sv} = \sqrt{2}\sigma_s'/R \quad (9)$$

Note that when $a = 0$, $\lim_{a \to 0} x_c = v_0 T/2$, $\lim_{a \to 0} \Delta\sigma_{sv}^2 = x_c^2/3$.

Let $\xi_{sv}^2 \triangleq (\xi_x + \delta x_v)^2 = \xi_x^2 + \delta x_v^2$, according to Eq.(3) and Eq.(9), the constraints on $x_c$ and $\sigma_s^2$ can be written as

$$\Delta\sigma_{sv}^2 \leq 4\varepsilon_1 \sigma_s^2/3 \quad (10a)$$

$$\Delta\sigma_{sv}^2 \leq \varepsilon_1' \xi_{\sigma^2} \quad (10b)$$

In the case that pixel size is $20''/pixel$, $\sigma_s = 2\,pixel$, $\varepsilon_1 = 0.1$, $\varepsilon'_1 = 2$, $T = 0.1s$ and $R = 50$ without jitter of motion and velocity aberration, Eq.(10a) suggests that the maximum velocity should be less than $0.14°/s$ while Eq.(10b) gives $v \leq 0.165°/s$ for in-orbit calibration. After motion compensation, the error from motion is limited or ignorable.

When $\Delta\sigma_{sv}^2 \geq \sigma_s^2$, Eq.(9) and Eq.(10a) show that the error from motion will dominate the process. It is proportional to $v_0^{3/2}$ under fixed exposure time. In this case, the star data set can only be used to output the attitude with an accuracy under a certain angular rate. Unfavorable error to correction will be introduced if the smeared star spot is used for in-orbit calibration. Once the rate exceeds a certain point, take $5°/s$ for example, the star spot is elongated so much that only the brightest star can be used to output the attitude with an error far beyond CRLB under static condition. To obtain a lower CRLB position error, both TDI and variable frame rate technology have been employed on star trackers[1-2]. Note that $x_c = 0.5$ when TDI technology is carried out, then CRLB could be written as $4\sqrt{\pi}\sigma_n'^2(\sigma_s^2 + 1/12)^{3/2}/\kappa^2 T^2$. Since $\sigma_s^2 \gg 1/12$, that is why implementation of TDI can obtain an accuracy virtually the same as static condition[1]. Variable frame rate is another effective approach to adjust the integration time under high velocity accordingly [2].

A drift in focal length is the **third** large error source. It results in an additional uncertainty about X and Y axes[1]. For the region far from the center of Field of View(FOV), this error is given by

$$\delta x_f = \frac{kd}{f}\delta f \quad (11)$$

where $k$ is the region position from center of FOV in unit of pixel on CCD, $d$ denotes the CCD pixel size in unit of $''/s$.

Various issues lead to a drift in focal length, such as temperature breathing effect and chromatic dispersion of optical lens[3-4]. Supposing the relationship between the temperature $\vartheta$ and focal length is approximate linear, It can be given by $f = f_0 + f_{\vartheta_0}(\vartheta - \vartheta_0) + \varsigma, f_\vartheta = \partial f/\partial\vartheta$  $\varsigma$ is the random error with $E[\varsigma] = 0, D[\varsigma] = \sigma_\varsigma^2$ when $\vartheta = \vartheta_0$. $\varsigma$ denotes the error sources such as color dispersion and the high order Taylor approximation. When temperature continues changing or chaos along the orbit[3,4,11], drift in focal length also causes another expansion of scale $\sigma_s$. This error is given by

$$\delta\sigma_f = \Delta\beta L = \frac{-\delta f}{\Delta f}\sigma_s \quad (12)$$

where $\Delta f$ is the defocus offset from the real focal length of lens.

For the defocused star spot, $\sigma_s$ can also be written as $\sigma_s = md, m > 0.5$ with $\delta\sigma_f^2 \approx 2\sigma_f \cdot \delta\sigma_f$. According to Eq.(3) and the focal length model, the constraints on $x_c$ and $\sigma_s^2$ can be written as

$$\frac{k}{f}(f_\vartheta^2 \text{var}(\tau) + \sigma_\varsigma^2)^{1/2} \leq \varepsilon_2\sqrt{2}m/R \quad (13a)$$

$$\frac{2\sigma_s^2}{\Delta f}(f_\vartheta^2 \text{var}(\tau) + \sigma_\varsigma^2)^{1/2} \leq \varepsilon'_2\xi_{\sigma^2} \quad (13b)$$

In the case that $\sigma_s = 2\,pixel$, $\varepsilon_2 = 0.5$, $\varepsilon'_2 = 0.3$, $k = 512$, $\Delta f = 5\times10^3\,\mu m$, $f = 5\times10^4\,\mu m$ and $R = 50$ under static condition, Eq.(17a) gives $\delta f \leq 2.76\mu m$ and Eq.(17b) gives $\delta f \leq 34.7\mu m$. After comparison with the error budget on motion, this study suggests that the drift of focal length might be the first dominant troublesome issue when star tracker is operating under static condition. The error from angular rate ranks the first place when star tracker is working under dynamic condition. Together with residual spatial error of FOV, It is branded as Low Frequency Error(LFE) in SED36 and HYDRA where LFE was demonstrated uncompensated[3]. Since the temperature breathing effect is more difficult to compensate, how to stabilize the temperature, improve structure of the thermal-mechanical component and reduce chromatic dispersion are big issues to solve. Eq.(10a) and Eq.(11) also indicates that large pixel size and small FOV is helpful to obtain a lower position error. It suggests that the orbit segmentation calibration technology might be a permitting way to limit the temperature effect. It also proves the advantage of local region correction technology[1-3]. It should be noted that the reason for local region calibration here is more because of drift in focal length other than optical aberration since the latter can be compensated while the former can only be depressed.

Note that each star acquired in all FOV will engage in outputting the attitude and each guide star in on-board star catalogue has possibility to be involved in in-orbit calibration. Those jitter effects, such as variable variance of noise, instable photo response non uniformity, color dispersion of star spot, instable intensity and so on are the **fourth** error contributors to the final accuracy. When CCD noises, intensity of star spot are changing in time, the jitter error on scale and intensity could be drawn as

$$\delta\sigma_j^2 \triangleq \Delta\xi_{\sigma^2} = \frac{\xi_{\sigma^2}}{R} - \xi_{\sigma^2}\frac{\Delta\sigma_n'}{\sigma_n'} \quad (14a)$$

$$\delta I_j \triangleq \Delta\xi_{I_0} = \frac{\xi_{I_0}}{R} + \xi_{I_0}\frac{\Delta\sigma_n'}{\sigma_n'} \quad (14b)$$

Since $\sigma_s = \beta L$ and $\Delta f/f \ll \Delta\lambda/\lambda$, another jitter effect from $\lambda$ is defined as $\delta\sigma_\lambda = \beta L(\Delta f/f + \Delta\lambda/\lambda)$. Incorporating $\delta\sigma_\lambda$ into Eq.(14a) and Eq.(14b) yields

$$\delta\Phi = \left[(\frac{\Delta\sigma_n'}{\sigma_n'})^2 + \frac{1}{R^2} + (\frac{\Delta\lambda}{\lambda})^2\right]^{1/2}\xi_\Phi \leq \varepsilon_\Phi\xi_\Phi \quad (15)$$

Note that an error on intensity and scale could introduce another error in position. Incorporating the jitter effect in Eq.(15) into position error, the total jitter error on position should be defined as

$$\delta x_j \triangleq \frac{\sqrt{2}\sigma_s}{R^2}(\frac{\partial R}{\partial\sigma_n'}\Delta\sigma_n' + \frac{\partial R}{\partial I_0}\delta I_j) + \frac{\sqrt{2}\delta\sigma_j}{R} + \frac{\sqrt{2}\delta\sigma_\lambda}{R} \quad (16)$$

Eq.(16) suggest the performance of PRNU and noise consistency of CCD devices should be stable to comply to the constraints in Eq.(3). The restrictions on any other influential factor could be derived in the same way. Set an appropriate value of $\varepsilon_\Phi$ in Eq.(15), the value ranges of those parameters are determined for in-orbit calibration before launch.

Note that an error scale from motion and drift in focal length could introduce another error in position. Then the final total position error from motion, drift of focal length and jitter effects can be given by

$$\delta x^2 \triangleq \delta x_v^2 + \delta x_f^2 + \delta x_j^2 = \frac{5\Delta\sigma_{sv}^2}{R^2} + \frac{k^2 d^2}{f^2}(f_\vartheta^2\text{var}(\vartheta) + \sigma_\varsigma^2)$$
$$+ \frac{4\sigma_s^2}{\Delta f R^2}(f_\vartheta^2\text{var}(\tau) + \sigma_\varsigma^2)^{1/2} + [1 + (\frac{\Delta\sigma_n'}{\sigma_n'})^2 + \frac{\zeta_{I_0}^2}{R^2} + (\frac{\Delta\lambda}{\lambda})^2$$
$$+ \frac{2\zeta_{\sigma^2}}{\sqrt{3}R}]\xi_x^2 \leq \zeta_x^2\xi_x^2 \quad (17)$$

where

$$\frac{\Delta\sigma_{sv}^2}{\xi_{\sigma^2}} + \frac{2\sigma_s^2}{\Delta f\xi_{\sigma^2}}(f_\vartheta^2\text{var}(\vartheta) + \sigma_\varsigma^2)^{1/2} + \left[1 + (\frac{\Delta\sigma_n'}{\sigma_n'})^2 + \frac{1}{R^2} + (\frac{\Delta\lambda}{\lambda})^2\right]^{1/2}$$
$$\triangleq \zeta_{\sigma^2}, \left[1 + (\frac{\Delta\sigma_n'}{\sigma_n'})^2 + \frac{1}{R^2}\right]^{1/2} \triangleq \zeta_{I_0}$$

Those functional error budgets demonstrate valuable features on device selection and system design and could save countless simulations before launch. As CRLB marks the margin for any variance or bias error, the error from those influential error sources could be very limited providing $\zeta_x$ in Eq.(17) is small enough. After precise compensation for some bias errors such as spatial error, the attitude accuracy after in-orbit correction on star tracker would be expected to approach the theoretical CRLB limits of CCD devices.

This work was supported by National Science Foundation (NSF) under grant 61174004, 61302117 and 2013 Annual National Major Instrument and Equipment Development Projects under grant 2013YQ310799.

**References**
1. R. W. H. van Bezooijen, Proc. SPIE 4850, 108-121(2003)
2. D. L. Michaels and J. F. Speed, Proc. SPIE 5430, 43-52(2004)
3. L. Blarre, N. Perrimon, D. Piot, et al., Proc. The 7th International ESA Conference on Guidance, Navigation&Control Systems, 2008.
4. C. C. Liebe, IEEE Trans. Aerosp. Electron. Syst. 38(2), 587-599(2002).
5. K. A. Winick, J. Opt. Soc. Am. A 3(11), 1809-1815(1986).
6. H. Cramér, *Mathematical Methods of Statistics*(Princeton University Press,1999).
7. C. R. Rao, *Breakthroughs in statistics*(Springer Series in Statistics, 1992).
8. C. Jutten and V. Vigneron, Lecture Notes in Computer Science 3195, 168-176(2004).
9. J. Zhang, Y. C. Hao, Y. Long, D. Liu, Acta Optica Sinica 35(8), 0825001(2015).
10. L. Lindegren, ISSI Scientific Report Series 9, 299-311(2013).
11. M. Gai, D. Carollo, M. Delbò, et al., Astronomy & Astrophysics 367(1), 362-370(2001).
12. R. A. Mendez, J. F. Silva and R. Lobos, Publ. Astron. Soc. Of the Pacific 125(927), 580–594(2013).
13. J. Arines and J. Ares, Opt. Lett. 27, 497 (2002).
14. M. A. Samaan, T. C. Pollock and J. L. Junkins, Journal of the Astronautical Sciences 50(1), 113(2002).
15. T. Sun, F. Xing, Z. You, et al., Optics express 21(17), 20096-20110(2013).
16. R. C. Gonzalez and R. E. Woods, *Digital image processing*(Pearson Education, 2009).